\documentstyle[amssymb,aps,12pt]{revtex}
\begin{document}
\draft
\author{Sergio De Filippo}
\address{defilippo@sa.infn.it}
\author{Filippo Maimone}
\address{maimone@sa.infn.it}
\address{Dipartimento di Fisica ''E.R. Caianiello'', Universit\`{a} di Salerno\\
Via Allende I-84081 Baronissi (SA) ITALY\\
and Unit\`{a} I.N.F.M., I.N.F.N., Salerno.}
\date{\today}
\title{Relativistic generalizations of gravity-induced localization models}
\maketitle

\begin{abstract}
Nonunitary versions of Newtonian gravity leading to wavefunction
localization admit natural special-relativistic generalizations. They
include the first consistent relativistic localization models. At variance
with the unified model of localization and gravity, the purely localizing
version requires negative energy fields, which however are less harmful than
usual and can be used to build ultraviolet-finite theories.
\end{abstract}


\newpage

Despite many claims according to which environment induced decoherence has
solved the measurement problem in quantum mechanics, the issue is not yet
settled\cite{Adler}. One of the proposals to cope with the measurement
problem is the modification of the evolution law in such a way to get the
emergence of classicality even in closed systems\cite
{GRW1,GPR,GRW2,Pearle1,Pearle2,Bassi}. Of course the modified dynamics must
comply with strict constraints, imposed by the huge amount of experimental
data consistent with the ordinary unitary, linear and deterministic
evolution law generated by the Hamiltonian operator. Although it was shown
that such constraints can be met by adding nonlinear stochastic terms to the
ordinary Schroedinger equation, that was achieved at some expense. First,
the proposed models require the introduction of phenomenological constants,
which should be fitted by future experiments. Secondly, as observed by John
Bell, who considered the main idea behind these models as a viable one, the
special role assigned to position requires a smearing on space, which makes
it quite problematic to find relativistic generalizations\cite
{Bell,Pearle2,Bassi}.

On the other hand the analysis of the possibility that the localization of
macroscopic bodies is an unavoidable effect of gravity has a long history
\cite{Karolyhazy,Penrose}. That idea led to the introduction of localization
models inspired by gravity, with the unattained aim of getting rid of the
mentioned free parameters\cite{Diosi,GGR}. It should also be mentioned that
a strong support to the idea that gravity may imply a nonunitary
generalization of quantum dynamics came from the emergence of the
information loss paradox within black hole physics\cite{Hawking2,Preskill}.

In some recent papers it was shown that suitable nonunitary modifications of
Newtonian gravity lead to localization models without any free parameter\cite
{Defilippo1,Defilippo2,Defilippo3}. While for previous nonunitary models
inspired by black hole dynamics the basic idea is to have the given system
interacting with a ''hidden system'' with ''no energy of its own and
therefore... not... available as either a net source or a sink of energy'' 
\cite{Unruh}, in the present models energy conservation is granted by the
''hidden system'' being a copy of the physical system, coupled to it only by
gravity, and constrained in its same state and then with its same energy.
The unitary dynamics and the states referred to the doubled operator algebra
are what we call meta-dynamics and meta-states, while, by tracing out the
hidden degrees of freedom, we get the non-unitary dynamics of the physical
states. Pure physical states correspond then to meta-states without
entanglement between physical and hidden degrees of freedom.

The hint that gravity may induce a nonunitary evolution came long ago even
from the perturbative analysis of Einstein gravity leading to the emergence
of higher order theories, which however are either nonunitary or plagued by
ghosts\cite{Dewitt,Stelle}. An optimistic conclusion is that ''the S matrix
will be nearly unitary''\cite{Dewitt}. In Ref. \cite{HawkingHertog} a remedy
for the ghost problem, leding to a nonunitary theory, was suggested by a
redefinition of the Euclidean path integral. A different approach in real
space-time was proposed in Refs. \cite{Defilippo5,DefMaim}, thus avoiding
analytical continuation, which amounts to a tricky operation outside the
realm of a fixed flat geometry. As in Ref. \cite{HawkingHertog}, classical
instability is cured at the expense of unitarity and the ensuing theory
singles out one of the mentioned modifications of Newtonian gravity as its
Newtonian limit. Of course the fully relativistic model may present the
usual problems ensuing from the consideration of a general covariant theory
of gravity within a quantum context.

In this letter we want to prove that there are no fundamental obstructions
to the building of relativistic localization models, by showing that the
mentioned nonrelativistic models have natural special-relativistic
generalizations, leading to the first well-defined localization models, both
relativistic and without free parameters.

In particular the analysis of the possible relativistic extensions sheds
some light on the ubiquitous presence of a divergent injection of energy in
the previous attempts\cite{Pearle2,Bassi}. Within the field theoretic
setting of the relativistic models presented here an uncontrollable energy
injection may occur only in the presence of negative energy fields. On the
other hand such fields are unavoidable within our approach only if one
requires that the localizing interaction averages out to zero. On the
contrary, if one accepts that the localizing interaction includes an average
effect, which in the nonrelativistic limit corresponds to the ordinary
Newtonian interaction, negative energy fields can be avoided.

To be specific, let $H_{0}[\psi ^{\dagger },\psi ]$ be the second quantized
form of an ordinary matter Hamiltonian in the absence of gravity. To define
the nonunitary Newtonian limit of the general covariant model\cite{DefMaim},
we introduce a (meta-)matter algebra that is the product of two equivalent
copies of the observable matter algebra, respectively generated by the $\psi
^{\dagger },\psi $ and $\tilde{\psi}^{\dagger },\tilde{\psi}$ operators and
a meta-Hamiltonian 
\[
H_{G}=H_{0}[\psi ^{\dagger },\psi ]+H_{0}[\tilde{\psi}^{\dagger },\tilde{\psi%
}] 
\]
\begin{eqnarray}
&&-\frac{G}{2}\sum_{j,k}m_{j}m_{k}\int dxdy\frac{\psi _{j}^{\dagger }(x)\psi
_{j}(x)\tilde{\psi}_{k}^{\dagger }(y)\tilde{\psi}_{k}(y)}{|x-y|}  \nonumber
\\
&&-\frac{G}{4}\sum_{j,k}m_{j}m_{k}\int dxdy\frac{\psi _{j}^{\dagger }(x)\psi
_{j}(x)\psi _{k}^{\dagger }(y)\psi _{k}(y)}{|x-y|}  \nonumber \\
&&-\frac{G}{4}\sum_{j,k}m_{j}m_{k}\int dxdy\frac{\tilde{\psi}_{j}^{\dagger
}(x)\tilde{\psi}_{j}(x)\tilde{\psi}_{k}^{\dagger }(y)\tilde{\psi}_{k}(y)}{%
|x-y|},  \label{newtonlimit}
\end{eqnarray}
acting on the product $F_{\psi }\otimes F_{\tilde{\psi}}$ of the Fock spaces
of the $\psi $ and $\tilde{\psi}$ operators. Here two couples of meta-matter
operators $\psi _{j}^{\dagger },\psi _{j}$ and $\tilde{\psi}_{j}^{\dagger },%
\tilde{\psi}_{j}$ \ appear for every particle species and spin component,
while $m_{j}$ is the mass of the $j$-th particle species. The $\tilde{\psi}%
_{j}$ operators obey the same statistics as the corresponding operators $%
\psi _{j}$, while $[\psi ,\tilde{\psi}]$ $_{-}=[\psi ,\tilde{\psi}^{\dagger
}]_{-}=0$.

The meta-state space $S$ is defined by a symmetry constraint as the subspace
of $F_{\psi }\otimes F_{\tilde{\psi}}$ including the meta-states obtained
from the vacuum $\left| \left| 0\right\rangle \right\rangle =\left|
0\right\rangle _{\psi }\otimes \left| 0\right\rangle _{\tilde{\psi}}$ by
applying operators built in terms of the products $\psi _{j}^{\dagger }(x)%
\tilde{\psi}_{j}^{\dagger }(y)$ and symmetrical with respect to the
interchange $\psi ^{\dagger }\leftrightarrow \tilde{\psi}^{\dagger }$,
which, then, have the same number of $\psi $ and $\tilde{\psi}$
meta-particles of each species. As the observable algebra is identified with
the $\psi $ operator algebra, expectation values can be evaluated by
preliminarily tracing out the $\tilde{\psi}$ operators.

It was shown that the ensuing non-unitary dynamics, while embodying the
ordinary Newton interaction, gives rise to a dynamical localization that is
compatible both with the wavelike behavior of microscopic particles and the
emergence of classicality for macroscopic bodies\cite
{Defilippo1,Defilippo2,Defilippo3,DefMaim,DefMaimRob}.

In an interaction representation, where the free meta-Hamiltonian is $%
H_{0}[\psi ^{\dagger },\psi ]+H_{0}[\tilde{\psi}^{\dagger },\tilde{\psi}]$,
if for simplicity we refer to one particle species only and $\psi ^{\dagger
}\psi $ denotes a quadratic scalar expression, the time evolution of a
generic meta-state $\left| \left| \tilde{\Phi}(0)\right\rangle \right\rangle 
$\ is represented by 
\begin{eqnarray}
\left| \left| \tilde{\Phi}(t)\right\rangle \right\rangle &=&{\it T}\exp 
\frac{i}{\hslash }Gm^{2}\int dt\int dxdy[\frac{\psi ^{\dagger }(x)\psi
(x)\psi ^{\dagger }(y)\psi (y)}{4|x-y|}+\frac{\tilde{\psi}^{\dagger }(x)%
\tilde{\psi}(x)\tilde{\psi}^{\dagger }(y)\tilde{\psi}(y)}{4|x-y|}  \nonumber
\\
&&+\frac{\psi ^{\dagger }(x,t)\psi (x,t)\tilde{\psi}^{\dagger }(y,t)\tilde{%
\psi}(y,t)}{2|x-y|}]\left| \left| \tilde{\Phi}(0)\right\rangle \right\rangle
\equiv U(t)\left| \left| \tilde{\Phi}(0)\right\rangle \right\rangle
\label{evolvedmetastate}
\end{eqnarray}
Then, by a Stratonovich-Hubbard transformation\cite{Negele}, we can rewrite $%
U(t)$ as a functional integral over an auxiliary real scalar field $\varphi $%
: 
\begin{eqnarray}
U(t) &=&\int {\it D}\left[ \varphi \right] \exp \frac{ic^{2}}{2\hslash }\int
dtdx\left[ \varphi \nabla ^{2}\varphi \right]  \nonumber \\
&&{\it T}\exp \left[ -i\frac{mc}{\hslash }\sqrt{2\pi G}\int dtdx\varphi
(x,t)\left( \psi ^{\dagger }(x,t)\psi (x,t)+\tilde{\psi}^{\dagger }(x,t)%
\tilde{\psi}(x,t)\right) \right] .  \label{stratonovich}
\end{eqnarray}
This allows, in particular, to obtain, by tracing out the hidden degrees of
freedom, an expression for the physical state $\rho _{Ph}$ evolving from the
generic pure state $\left| \Phi (0)\right\rangle _{\psi }$, which can be
taken as an alternative definition of the model, independent of any
reference to the hidden degrees of freedom\cite{Defilippo3,DefMaim}: 
\[
\rho _{Ph}(t)=\int {\it D}\left[ \varphi ,\varphi ^{\prime }\right] \exp 
\frac{ic^{2}}{2\hslash }\int dtdx\left[ \varphi \nabla ^{2}\varphi -\varphi
^{\prime }\nabla ^{2}\varphi ^{\prime }\right] 
\]
\begin{eqnarray}
&&_{\psi }\left\langle \Phi (0)\right| {\it T}^{-1}\exp \left[ i\frac{mc}{%
\hslash }\sqrt{2\pi G}\int dtdx\varphi ^{\prime }\psi ^{\dagger }\psi \right]
\nonumber \\
&&{\it T}\exp \left[ -i\frac{mc}{\hslash }\sqrt{2\pi G}\int dtdx\varphi \psi
^{\dagger }\psi \right] \left| \Phi (0)\right\rangle _{\psi }  \nonumber \\
&&{\it T}\exp \left[ -i\frac{mc}{\hslash }\sqrt{2\pi G}\int dtdx\varphi \psi
^{\dagger }\psi \right] \left| \Phi (0)\right\rangle _{\psi }  \nonumber \\
&&_{\psi }\left\langle \Phi (0)\right| {\it T}^{-1}\exp \left[ i\frac{mc}{%
\hslash }\sqrt{2\pi G}\int dtdx\varphi ^{\prime }\psi ^{\dagger }\psi \right]
.  \label{alternative}
\end{eqnarray}

The announced relativistic model is obtained by the immediate generalization
of the equation above corresponding to the replacement of the matter fields
with their relativistic generalization and of the Laplacian with the
d'Alambertian operator. The same replacement transforms Eq. (\ref
{stratonovich}) into a mixed operator and path integral expression for a
theory with meta-matter interacting with a quantum neutral scalar field by a
Yukawa interaction. The ensuing theory is of course a well defined
renormalizable field theory without any instability leading to an
uncontrollable increase of the matter energy.

If one assumes that the ensuing relativistic model is a real improvement on
its Newtonian limit, one has to see if using the latter is consistent at
all. In order to do that, consider that the Newtonian model gives a
localization length $\Lambda \sim (\hslash ^{2}R^{3}/GM^{3})^{1/4}$ for a
body whose linear dimension is $R$ and whose mass $M$ is above the
threshold, which for ordinary densities is $\sim 10^{11}$ proton masses ($%
m_{p}$)\cite{Defilippo1,DefMaim}. The localization process implies a
localization energy $E_{\Lambda }\sim \hslash ^{2}/(M\Lambda ^{2})\sim
\hslash G^{1/2}\rho ^{1/2}$, depending only on the body density $\rho $,
which, for ordinary densities is $E_{\Lambda }\sim 10^{-20}eV$. This process
takes a time $T_{G}\sim 10^{20}(M/m_{p})^{-5/3}s$\cite{Defilippo3,DefMaim}
and consists of the transformation of potential into kinetic meta-energy,
corresponding to twice the physical kinetic energy.

To estimate the radiated energy in the relativistic model, consider that the
bound metastates, corresponding to localized states, are small oscillations
in a potential $U(r)\sim (GM^{2}/R^{3})r^{2}$, namely with a frequency $%
\omega \sim \sqrt{GM/R^{3}}$. The corresponding classical radiating power
for the $n$th harmonic frequency is easily seen to be $w_{n}=(n\omega
)^{2}GM_{n}^{2}/c$, where $\rho (x,t)\equiv \sum_{n}\rho _{n}(x)\exp
(in\omega t)$, $M_{n}\equiv \int dx\rho _{n}(x)$. For ordinary densities and 
$M=10^{12}m_{p}$, just above the threshold, one gets a total radiating power 
$w\lesssim 10^{-37}eV/s$, which in the localization time $T_{G}\sim 1s$
amounts to an irrelevant fraction ($\sim 10^{-17}$) of the localization
energy $E_{\Lambda }$. This means that, in order to estimate relativistic
corrections, it makes sense just to replace in Eq. (\ref{newtonlimit}) the
instantaneous interactions with the ones mediated by the retarded propagator.

Also in Pearle's proposal\cite{Pearle3,Bassi} matter is coupled to a scalar
field by a Yukawa interaction. The main differences consist in the field
being massive and in the fact that here it is coupled to a hidden copy of
matter, whereas in Pearle's model it is coupled to a classical stochastic
field, whose c-number character leads to an infinite growth rate of the
energy of the scalar field\cite{Bassi}. Moreover, while our aim was to build
a unified model of localization and gravity, the interaction introduced in
Pearle's model is meant to produce localization only.

If one wanted to introduce a scalar field leading only to dynamical
localization, without an average ensuing force, then one would be forced to
take a negative energy field, namely an evolution operator 
\begin{eqnarray}
U_{0}(t) &=&\int {\it D}\left[ \varphi \right] \exp \frac{i}{2\hslash }\int
dtdx\left[ -c^{2}\varphi \nabla ^{2}\varphi +\varphi \partial
_{t}^{2}\varphi \right]  \nonumber \\
&&{\it T}\exp \left[ -i\frac{mc}{\hslash }\sqrt{2\pi G}\int dtdx\varphi
(x,t)\left( \psi ^{\dagger }(x,t)\psi (x,t)-\tilde{\psi}^{\dagger }(x,t)%
\tilde{\psi}(x,t)\right) \right] .  \label{pureloc}
\end{eqnarray}
In fact, if one integrates out the scalar field, in analogy with the
Feynman's elimination of the electromagnetic potential\cite{FeynmanHibbs}
and one takes the $c\rightarrow \infty $ limit, one gets a Hamiltonian like
the one in Eq. (\ref{newtonlimit}), but for the replacement $G\rightarrow -G$
in the last two terms. By paraphrasing the analysis performed in Refs.\cite
{Defilippo1,DefMaim}, one sees that, once the symmetry constraint is
considered, no net force survives and that the localization properties are
exactly the same as for the original model, since they depend only on the
interaction between physical and hidden degrees of freedom. It should be
remarked that if in Eq. (\ref{pureloc}) one replaces the scalar field with a
positive energy one, by changing the sign in the exponent of the first
exponential, one gets a model still with a vanishing net force, but without
localization properties, as the interaction between physical and hidden
degrees of freedom is repulsive.

Although our aim was the introduction of a well-defined relativistic theory,
on a heuristic level one can introduce another relativistic model where both
positive and negative energy fields are present, with the further bonus of
Pauli Villars-like cancellations, like in the general covariant theory\cite
{Stelle,Defilippo5,DefMaim}. In fact, if we consider a relativistic action 
\begin{eqnarray}
{\cal A} &=&{\cal A}_{0}[\psi ^{\dagger },\psi ]+{\cal A}_{0}[\tilde{\psi}%
^{\dagger },\tilde{\psi}]  \nonumber \\
&&+\frac{1}{2}\int dtdx[c^{2}\varphi _{1}\nabla ^{2}\varphi _{1}-\varphi
_{1}\partial _{t}^{2}\varphi _{1}-c^{2}\varphi _{2}\nabla ^{2}\varphi
_{2}+\varphi _{2}\partial _{t}^{2}\varphi _{2}  \nonumber \\
&&-2mc\sqrt{2\pi G}(\psi ^{\dagger }\psi +\tilde{\psi}^{\dagger },\tilde{\psi%
})\varphi _{1}-2mc\sqrt{2\pi G}(\psi ^{\dagger }\psi -\tilde{\psi}^{\dagger
},\tilde{\psi})\varphi _{2}]  \label{finite}
\end{eqnarray}
where ${\cal A}_{0}[\psi ^{\dagger },\psi ]$ is the ordinary relativistic
matter action, its Newtonian limit is obtained by the elimination of the
second and the third term and the replacement of $G$ with $2G$ in Eq. (\ref
{newtonlimit}). This nonrelativistic model is qualitatively equivalent to
the Newtonian limit of the general covariant theory, apart from little
quantitative changes in the localization properties due to the doubling of
the localizing interaction. As to the relativistic model (\ref{finite}), it
is remarkable that it contains no new ultraviolet divergences with respect
to the ones already present in the traditional theory with action ${\cal A}%
_{0}[\psi ^{\dagger },\psi ]$, as there is a complete cancellation of all
self-energy and vertex graphs corresponding to the interaction of
meta-matter with the scalar fields, due to the difference in sign of their
propagators.

In conclusion some remarks are in order.

First it should be added that in our models (the avoidable) negative energy
fields are less harmful than expected, since their average values are
constrained to vanish, which makes such models stable, at least within a
naive classical analysis, like it happens in the general covariant theory
\cite{Defilippo5,DefMaim}. This follows from the fact that, in order for the
evolution to be compatible with the symmetry constraint, one has to
generalize the latter by replacing the symmetry transformation exchanging
physical and hidden degrees of freedom with 
\begin{equation}
\psi \rightarrow \tilde{\psi},\;\;\tilde{\psi}\rightarrow \psi
,\;\;\;\varphi \rightarrow -\varphi ,\;  \label{symmetry}
\end{equation}
where $\varphi $ is the negative energy field, in analogy to Eq. (12) in Ref.
\cite{DefMaim}.

Secondly we should stress that, even though the obstruction to the
formulation of consistent relativistic localization models is removed within
a unified theory of localization and gravity, this does not mean that our
special relativistic extensions may include a relativistic theory of
gravity. In fact the Newtonian interaction is obtained starting from a
Yukawa interaction, while a relativistic theory of gravity should involve
the matter energy momentum tensor. However, the present results, together
with the observation that renouncing unitarity may tame the instabilities
inherent in higher order gravity\cite{Defilippo5,DefMaim}, appear to us to
be a rather compelling indication that a unified relativistic theory of
spontaneous localization and gravity may be easier to construct than a
unitary theory of gravity.

Finally one can look in principle for spontaneous localization models in
terms of a stochastic dynamics for pure states, which, when averaged, leads
to our nonunitary evolution of the density operator\cite
{GRW1,GPR,GRW2,Pearle1,Pearle2,Bassi}. Apart from the non-uniqueness of
stochastic realizations\cite{Pearle2}, stochastic models can certainly be
useful as computational tools. However the view advocated here considers
density operators as the fundamental characterization of the system state
and not just as a bookkeeping tool for statistical uncertainties. This point
of view, apart from possibly being relevant to the quantum foundations of
thermodynamics\cite{DefMaim}, avoids the ambiguities of the stochastic
viewpoint, where the expectation of a local observable depends on the choice
of a particular space-like surface in its entirety ( Ref.\cite{Bassi}, sect.
14.2). The fact that, in measurement processes, the apparent uniqueness of
the result seems to imply a real collapse is perhaps more an ontological
than a physical problem, and presumably, if one likes it, that can be
addressed by a variant of the Everett interpretation\cite{Everett}.

\end{document}